\documentclass[prb,superscriptaddress,twocolumn,aps,showpacs,preprintnumbers,amsmath,amssymb,floatfix]{revtex4}

\usepackage{graphicx}
\usepackage{wasysym}

\newcommand{\be}{\begin{equation}}
\newcommand{\ee}{\end{equation}}
\newcommand{\bea}{\begin{eqnarray}}
\newcommand{\eea}{\end{eqnarray}}

\begin{document}

\preprint{2D-sp-AM/INT}


\title{Wave function statistics at the symplectic 2D Anderson
  transition: bulk properties}

\author{A.~Mildenberger}%
  \affiliation{\mbox{Fakult\"at f\"ur Physik, Universit\"at Karlsruhe, 
    76128 Karlsruhe, Germany}} 
\author{F.~Evers}%
  \affiliation{Institut f\"ur Nanotechnologie, 
    Forschungszentrum Karlsruhe, 76021 Karlsruhe,
    Germany}
  \affiliation{\mbox{Institut f\"ur Theorie der Kondensierten Materie, Universit\"at Karlsruhe, 
    76128 Karlsruhe, Germany}} 

\date{\today}

\begin{abstract}
The wave function statistics at the Anderson transition in a two-dimensional
disordered electron gas with spin-orbit coupling is studied numerically. 
In addition to highly accurate exponents
($\alpha_0{=}2.172\pm 0.002, \tau_2{=}1.642\pm 0.004$),
we report three qualitative results. (i) The anomalous dimensions
are invariant under $q\rightarrow (1-q)$ which is
in agreement with a recent analytical prediction and supports the
universality hypothesis.
(ii) The multifractal spectrum is not parabolic and therefore differs
from behavior suspected, e.g., for (integer) quantum Hall transitions
in a fundamental way. (iii) The critical fixed point satisfies
conformal invariance.
\end{abstract}


\maketitle


Disordered electron systems that are confined to two spatial dimensions (2D)
cannot support a true metallic state because of 
{\it Anderson localization}.\cite{anderson58} The underlying physics
relates to an interference-enhanced return probability of quantum mechanical 
particles due to repeated backscattering
of the same (quenched) disorder configuration. 
There are exceptions to the rule, however. For instance, if spin-orbit
scattering exists, the return probability is not enhanced but even 
depleted and the metallic state survives.\cite{lee85} Universal properties of such
metals are described by the {\em symplectic} symmetry class of 
Gaussian random matrix theories. By increasing the disorder strength $W$, a
metal-insulator (i.e. {\em Anderson transition}) can be driven in these 
materials. Its universal properties have been 
studied intensively in the last two decades. 

One of the controversial questions in the late 1990s concerning the 
symplectic transition in 2D was about the numerical value of the
critical exponent $\nu$ that describes the divergence of the localization
length when the disorder approaches its critical value:
$\xi\sim|W-W_c|^{-\nu}$.
In recent work, Asada {\it et al.} have made a very
convincing case in favor of $\nu{=}2.75$
(overview in Table \ref{t1}) employing the SU(2) model.\cite{asada04a}
A work by Markos and Schweitzer\cite{markos05} comes 
to a similar conclusion, $\nu\approx
2.8\pm0.04$, within the {\em Ando model} 
and the debate is now settled. 

However, this latter work not only has helped to fix $\nu$, it also has
reemphasized that another important topic is still unresolved. 
Recall that the critical wave functions, $\Psi({\bf x})$
at the boundary between
insulator and metal obey a {\it multifractal statistics}.\cite{mirlin00} 
This implies that the moments 
\be
\label{e1}
\langle\!\langle |\Psi({\bf x})|^{2q}\rangle\!\rangle \sim L^{-d-\tau_q},
\qquad q\in \mathbb{R}
\ee
scale with system size $L$, introducing the exponent spectrum 
$\tau_{q}$. (The angular brackets denote a combined spatial and ensemble average.)
A precise numerical determination of 
$\tau_{q}$ has not been undertaken yet.
The numerical work presented in this letter is an attempt to close
this gap. 

There are several good reasons why one would like to scrutinize
the nature of $\tau_q$ more closely. For one thing, the wave function
statistics can be measured, in principle, and promising steps in this
direction were made not long ago.\cite{morgenstern03}

But also important questions concerning our conceptual
understanding of the localization-delocalization transition are
closely related to multifractality.
First, the analytic structure of $\tau_q$ 
is a specific characteristic of the critical field theory of the
transition describing scaling of the local density of states. For
example, it has
been proposed that the (integer) quantum Hall transition exhibits
reduced anomalous dimensions $\delta_q$,  
\be
 \tau_q = d(q-1) + \delta_q q (1-q),
 \label{e2}
\ee
with a special property: $\delta_q$ does not depend on $q$,
such that $\tau_q$ is parabolic and also invariant under
$q\rightarrow 1{-}q$.
Very recently, it has been predicted\cite{mirlin06a} -- based on exact results for the
nonlinear $\sigma$ model and invoking the universality hypothesis --
that this last symmetry is a general property of all
transitions belonging to 
the conventional Wigner-Dyson classes. That is 
\be
\delta_q = \delta_{1-q}
\label{e3}
\ee
should hold. A numerical verification beyond the framework of the
power law random banded matrix model has not been reported yet.
This would be an interesting test of universality, since it does
not only rely on comparing quantitative values for some few exponents 
-- which has been the usual procedure --
but rather refers to the analytic structure of an exponent spectrum. 
Note that Eq. (\ref{e3}) does not generally hold outside the conventional
symmetry classes.
The spin quantum Hall effect is an example for a
transition in a  nonstandard universality class, where  Eq. (\ref{e3}) 
is manifestly violated.\cite{evers02,mirlin02}

Second, lately it has become clear that near boundaries
multifractality differs from the bulk: flat interfaces support 
their  own ``surface'' spectrum $\tau^{\rm s}_q$; in the presence
of corners yet another spectrum is superimposed, etc.\cite{subramaniam06}
\onecolumngrid
\begin{center}
\begin{table}
\begin{tabular}{|c|c|c|c|c|c|c|c|}\hline\hline
model & method & $W_c$         & $\Lambda_c$       & $\alpha_0=2+\delta_0$ & $\delta_q$ & $\nu$ & reference\\\hline\hline
SU(2) & TM     &$5.953\pm0.001$&$1.843\pm0.0013$& & & $2.746\pm0.009$ &\onlinecite{asada04a} \\\hline
AM    & TM     &$5.838\pm0.007$&$1.87\pm0.02$ &&&$2.8\pm0.04$&\onlinecite{markos05}\\ 
    & MAt    &$5.838\pm0.007$&&$2.107\pm0.005$&$\delta_1{=}0.111$&&\onlinecite{markos05}\\ \hline
AM&MAt&$5.86\pm0.04$&&&$\delta_2=0.19\pm0.005$&$2.41\pm0.24$&\onlinecite{yakubo98}\\\hline
AM&wave-packet propagation&5.74&&&$\delta_2{=}0.15\pm0.02$&&\onlinecite{kawarabayashi96}\\\hline
AM&MAt&5.74&&$2.19\pm0.03$&$\delta_2{=}0.17\pm0.025$&&\onlinecite{schweitzer95}\\\hline
EZM&MAa& & & & $\delta_1{=}0.16\pm 0.02$  & & \onlinecite{evangelou90}\\
   &   & & & & $\delta_2{=}0.185\pm 0.01$ & & \onlinecite{chalker93}\\\hline
network model&&&$1.83\pm0.03$&&&$2.51\pm0.18$&\onlinecite{merkt98}\\\hline
\hline
\end{tabular}
\caption{\label{t1} Overview of results for 
  the symplectic transition in two dimensions.
  AM: Ando model\cite{ando89}; EZM:
  Evangelou-Ziman-model\cite{evangelou87}; MAt (MAa): multifractal
  analysis based on scaling of typical (average) amplitudes; 
  TM: transfer matrix; 
  SU(2): SU(2) model\cite{asada02};
  $\delta_q$: reduced anomalous dimension, see Eq. (\ref{e2}).
  Entries for the same model are in chronological order, starting with the
  latest work. }
\vspace*{-1cm}
\end{table}
\end{center}
\twocolumngrid
\noindent
Also, in principle, an edge could break a bulk symmetry and thus would not even
share the bulk universality class.
In fact, the unraveling of surface multifractality could lead to a
paradigmatic shift of our present understanding of 
critical wave function statistics. 
Clearly, a prerequisite for all this is a detailed
knowledge of bulk properties.  

Third, finally, a relation between $\delta_0$
and the ratio $\Lambda_c$ of width and localization length of 
quasi-1D strips exists:
\be
\label{e4}
    \Lambda_{c}=1/\pi\delta_0,
\ee
which is exact {\em if} the critical 2D fixed point
is {\em conformally invariant}.\cite{janssen94a}
It is believed that conformal invariance (CI) is a generic property of
localization-delocalization transitions in 2D.
For instance, it has been demonstrated to hold at the
integer quantum Hall transition.\cite{janssen99,klesse00}
Exceptions are not known so far, but Eq. (\ref{e4}) can be used 
as a test of CI. In this respect,
recent numerical results are alarming.
It is reported\cite{markos05} that $\delta_0{=}0.107\pm0.005$ 
and  $\Lambda_{c}{=}1.87\pm0.02$; thus the product
$\pi\Lambda_{c}\delta_{0} = 0.629 \pm 0.036$ would signal 
a {\em strong violation} of Eq. (\ref{e4}) and therefore absence of
CI.\cite{olddelta0}

In this Rapid Communication, we present a numerical high-precision
study of $\delta_q$ at the 2D-symplectic transition.
Our particular aim is to answer three qualitative questions.
(i) Is $\delta_q$ a constant, so $\tau_q$ is parabolic?
(ii) If not, does it obey the symmetry relation Eq. (\ref{e3})
confirming the universality hypothesis?
(iii) Is the fixed point conformally invariant?

Most earlier works analyzed
typical moments in small ensembles, where finite-size
effects make it difficult to obtain reliable error bars.
By contrast, we employ scaling of typical and average
moments in very large ensembles with big system sizes.
Errors can thus be reduced by almost an order of magnitude. 
In order to cross-check, we analyze the two most important microscopic 
models. Results thus obtained agree very well.  
Specifically, we find that $\delta_q$ is not a constant and the symmetry relation
(\ref{e3}) is satisfied.

On a quantitative level, we obtain
$\delta_2{=}0.180{\pm}0.002$ (both models), 
$\delta_0{=}0.173{\pm}0.003$ (Ando model), and   
$\delta_0{=}0.172{\pm}0.002$ (SU(2) model).
Together with Eq. (\ref{e4})
and the earlier result\cite{asada04a} $\Lambda_c{=}1.843$
we arrive at
$\pi\Lambda_c\delta_0=0.996{\pm}0.012$.
Thus numerical evidence is provided that the symplectic
fixed point obeys CI, in agreement with general expectations.


\paragraph*{Models:}
We consider a tight-binding Hamiltonian on a two dimensional square lattice
with nearest neighbor coupling
\be
  H = \sum_{i,\sigma} \epsilon_i^{\phantom{\dagger}} 
           c^\dagger_{i,\sigma} c^{\phantom{\dagger}}_{i,\sigma}
    + \sum_{\langle i,j \rangle, \sigma, \sigma'} 
         V_{i,\sigma;j,\sigma'}^{\phantom{\dagger}}
         c^\dagger_{i,\sigma} c^{\phantom{\dagger}}_{j,\sigma'} ,
\ee
where $c^\dagger_{i,\sigma}$ ($c^{\phantom{\dagger}}_{i,\sigma})$ denotes
a creation (annihilation) operator of an electron with spin $\sigma$
on site $i$.

In the {\it Ando model}\/\cite{ando89}, 
the on-site energies $\epsilon_i$ are taken independently 
from the interval $[-W/2,W/2]$ with a homogenous distribution. 
The hopping matrix $V_{i,\sigma;j,\sigma'}$
reflecting the spin-orbit coupling is chosen as
\be
  V_{i,\sigma;i+k,\sigma'} = 
    \left( V_0 \exp (i \theta_k \sigma_k) \right)_{\sigma,\sigma'}, \qquad
  k = x,y,
\ee
with $\sigma_x, \sigma_y$ denoting Pauli matrices and the 
parameters $V_0=1$ and $\theta_k= \pi/6$. 
We have 
determined the critical disorder strength independently via analysis of
the critical level statistis.\cite{scalingprocedure}
Our finding  $W_c{=}5.85{\pm}0.025$
agrees well with earlier work.\cite{markos05}

The second model, the {\it SU(2) model}, has been introduced by 
Asada, Slevin, and Ohtsuki \cite{asada02}.
In addition to the on-site energies $\epsilon_i$, 
now also the hopping matrix $V_{i,\sigma;j,\sigma'}$ is random. 
It is taken to be uniformly distributed over
the entire group SU(2) using the group invariant ({\it Haar}) measure.\cite{asada02}

$H$ is implemented on square $L \times L$-size lattices 
with periodic boundary conditions. For our numerical
diagonalization of the resulting $2L^2 \times 2L^2$ matrices we 
use an inverse iteration routine coupled with direct sparse solvers 
in order to obtain the eigenvalues and wave functions with energies closest 
to zero.\cite{arpack} (Cf. Ref. \onlinecite{evers01}.) 
\begin{figure}[t]
\includegraphics[width=0.9\columnwidth,clip]{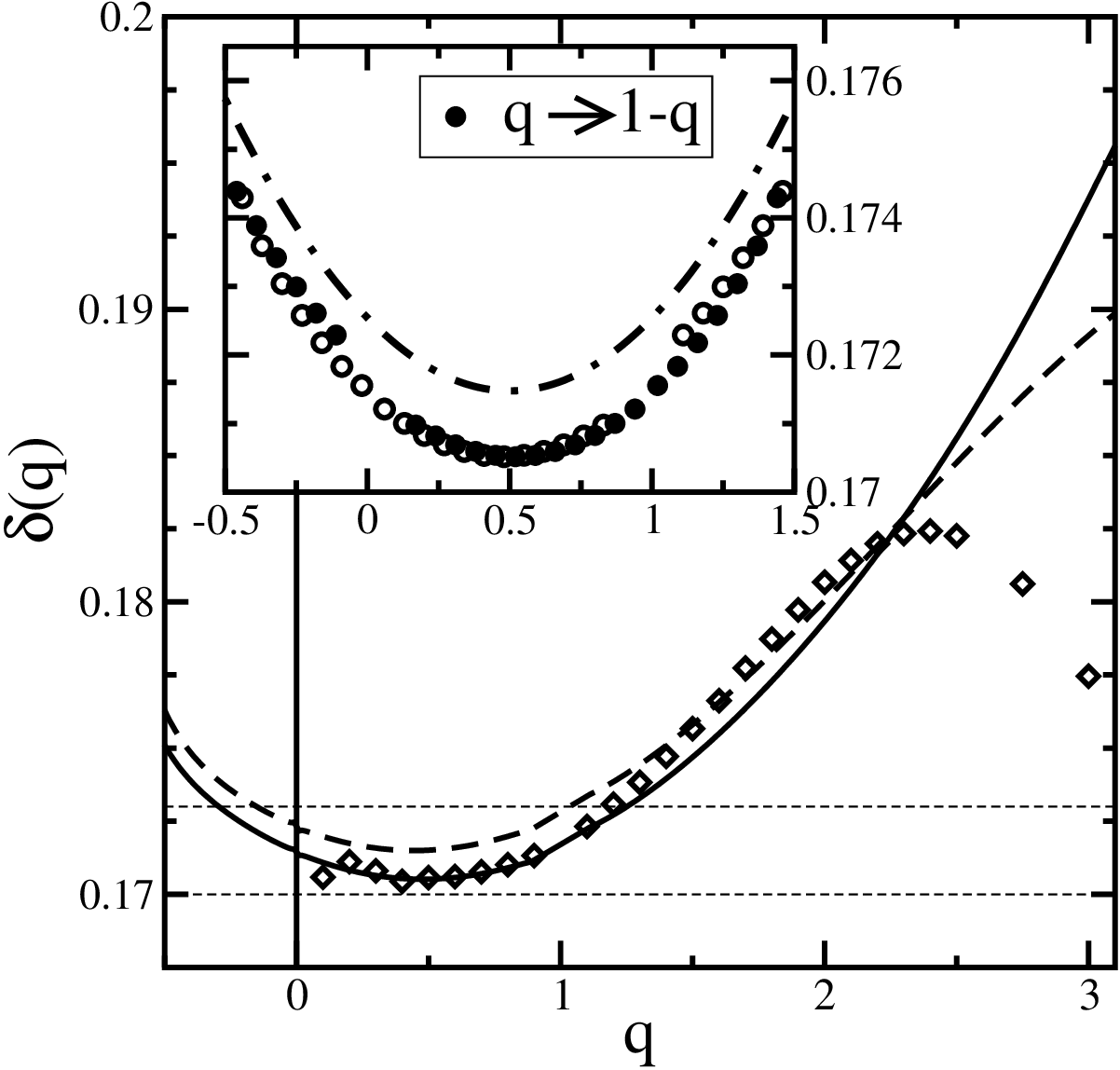}
\caption{Reduced anomalous dimension $\delta_{q}$  as defined in
  Eq. (\ref{e3}) for the Ando model (dashed, $W_c{=}5.84$)
  and the SU(2) model (solid, $W_c{=}5.953$). Additionally,
  anomalous dimensions $\tilde\delta_q$ obtained from 
  typical inverse participation ratios are shown ($\diamond$) for the latter
  model. 
  Dashed lines indicate
  the estimated error (2$\sigma$) in $\delta_0$. 
  Inset: blowup of the solid line behavior near $q{=}0.5$ 
  now represented by $\circ$. Data near $q{=}0$ and $q{=}1$ suffer from noise
  amplification (dividing by $q(1{-}q)$ in Eq. (\ref{e2})) and have
  therefore been omitted. Filled symbols ($\bullet$) show original trace
  after reflection at $q{=}0.5$. Dot-dashed line indicates parabolic fit
  (offset: $10^{-3}$) with $\delta_{1/2}{=}0.1705$ and
  curvature $\delta^{\prime\prime}_{1/2}{=}0.0043.$}
\label{f1}
\end{figure}


\paragraph*{Multifractal analysis:}
Our multifractal analysis proceeds by analyzing the scaling behavior of the 
average moments of wave function amplitudes, Eq. (\ref{e1}).

In order to analyze the critical behavior we take
the disorder value $W_c{=}5.84$ (for states at energy
zero being critical) in the Ando model. For the 
SU(2) model we employ $W_c{=}5.953$ in order to have
a mobility edge at energy $\epsilon{=}1$.\cite{asada04a}
The average (\ref{e1}) has been performed over an ensemble of
wave functions that have been calculated in systems of sizes
$L{=}16,24,32,48,64,96,128,192,256$ (the last two values were not used
in all cases). For each disorder realization $64$ wave functions closest
to the critical energy have been taken into account; all together
the number of wave functions in the ensemble is typically 
$4\times 10^7$ ($L{=}16$) to $3\times 10^5$ ($L{=}256$).

The exponents $\tau_q$ are readily extracted from a power-law fit
as suggested by Eq. (\ref{e1}).\cite{extracttauq}
In Fig. \ref{f1}
we plot the reduced dimensions
$\delta_q$ defined in (\ref{e2}) as obtained for both models.
It incorporates our three main results.

(i) We determine $\delta_0{=}0.172\pm 0.002$. The value satisfies
Eq. (\ref{e4}) and thus the consistency check on CI is positive.
The good accuracy
stems mainly from large statistics and the fact,
that finite-size corrections in the
SU(2) model turn out to be extremely small at $q\apprle 1.5$.
As can also be seen from Fig. \ref{f1}, the Ando model gives
a similar result.

(ii) The function $\delta_q$ satisfies the symmetry relation
Eq. (\ref{e3}). Thus the universality postulate is confirmed.
The inset of Fig. \ref{f1} shows that
part of the full curve $\delta_q$, for which numerical data are
available at both points, $q$ and its image $1-q$.
(The numerical procedure that we work with is limited to 
$q {\apprge}-1$; more negative values would require a coarse graining
in order to overcome the divergence of the moments (1) related to
zeros of the wave functions.)
A symmetric shape of the curve is clearly
displayed in the regime of best accuracy, $-0.5 \apprle q \apprle 1.5$. 
\begin{figure}[t]
\includegraphics[width=0.9\columnwidth,clip]{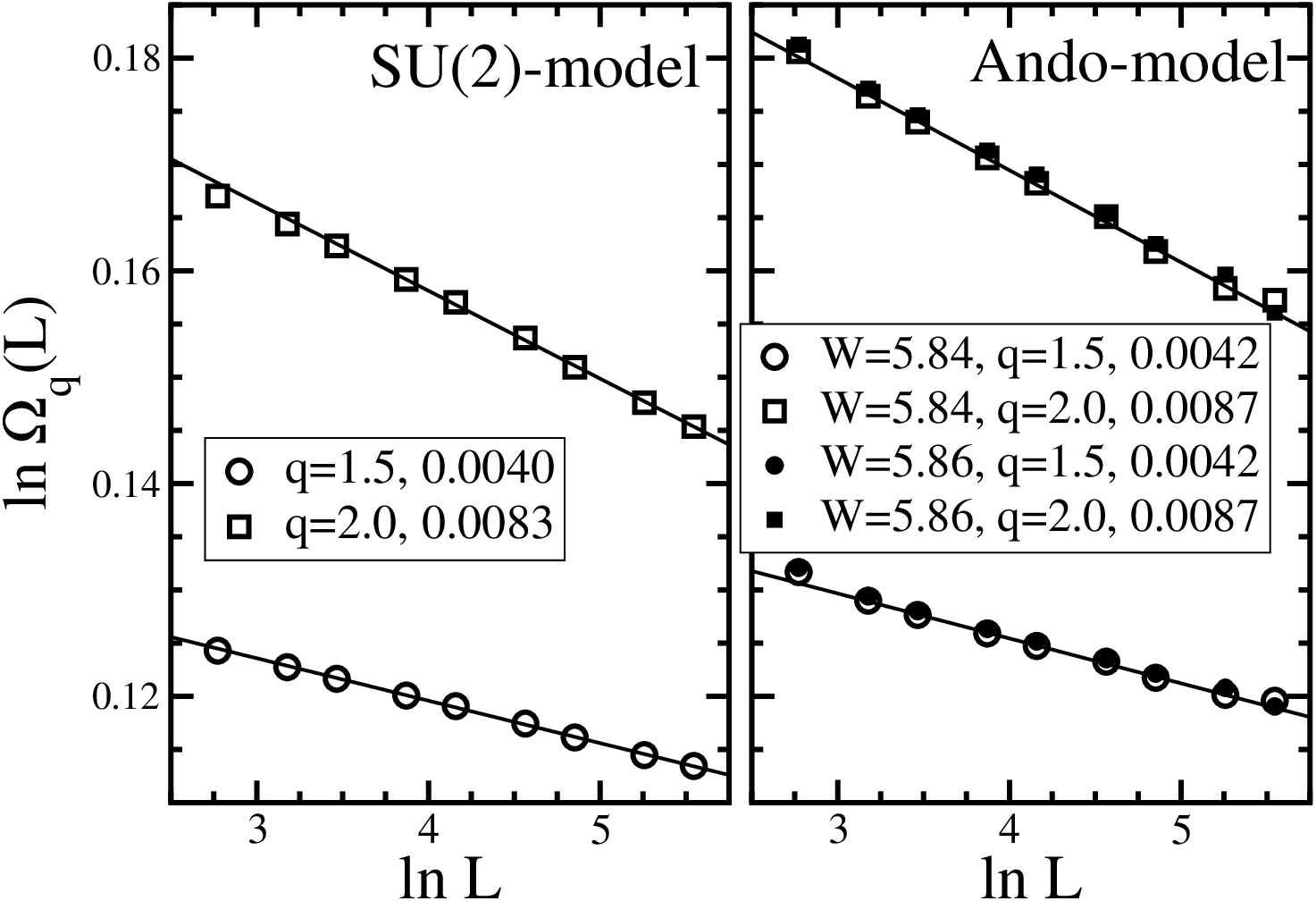}
\caption{Test function $\Omega_q(L)$ highlighting variability of $\delta_q$ with
$q$ for two $q$ values. Solid lines represent power law fits, with a 
  slope representing $\delta_q-\delta_{1/2}$; for values, see legend.
Slight deviations between models are due to larger
errors in finite-size extrapolation of Ando model.
For that model, results for two values of $W$ are given 
to illustrate that the uncertainty in $W_c$ is not a 
precision-limiting factor. 
Note that $\delta_{3/2}{-}\delta_{1/2}$ agrees well with curvature
$\delta^{\prime\prime}_{1/2}$ seen in Fig. \ref{f1}. 
}
\label{f2}
\end{figure}

(iii) The set of exponents $\delta_q$ does not reduce to
a constant, e.g., $\delta_q$ has a small but nonzero
curvature $\delta^{\prime\prime}_{1/2}$.
Detecting $\delta^{\prime\prime}_{1/2}$ requires high-precision data,
because the  numerical window is
limited to $q\apprle 2.0$. At larger values, (a) finite-$L$ effects
proliferate (in Ando model faster than in SU(2)), so deviations
between solid and dashed lines increase. And (b)
moments $\langle\!\langle |\Psi|^{2q}\rangle\!\rangle$ for large $q$ probe the tails of the
distribution function, so that typical values and averages differ from
each other. Then, error bars tend to become large due to
undersampling.\cite{evers01} The parting of the three
curves at $q\apprge 2$ visible in Fig. \ref{f1} is a consequence of
these effects.

As a sensitive test for variability of $\delta_q$ we investigate 
in  Fig. \ref{f2} the ratio
\be
   \Omega_q(L) = 
\Bigl[ \langle\!\langle |\Psi|^{2q} \rangle\!\rangle \, L^{d\,q} \Bigr]^{1/q(1-q)}
/ \Bigl[ \langle\!\langle |\Psi| \rangle\!\rangle \, L^{d/2} \Bigr]^{4}
\ee
encompassing only unprocessed data. It scales as
$
\Omega_q(L) {\sim} L^{-\delta_q {+} \delta_{1/2}}
$
and therefore any slope in $\ln \Omega$ signalizes that 
$\delta_q$ deviates from $\delta_{1/2}{=}0.1705\pm0.001$.
Data for $\Omega_{q}$ at $q{=}1.5,2.0$
are shown in Fig. \ref{f2}. It clearly exhibits a linear trace 
with the nonzero slope indicative of curvature in $\delta_q$.
Note that finite-size effects are 
very small, so that 
$\delta_q{-}\delta_{1/2}$ can be extracted with
good accuracy.  

\begin{figure}[t]
\includegraphics[width=0.8\columnwidth,clip]{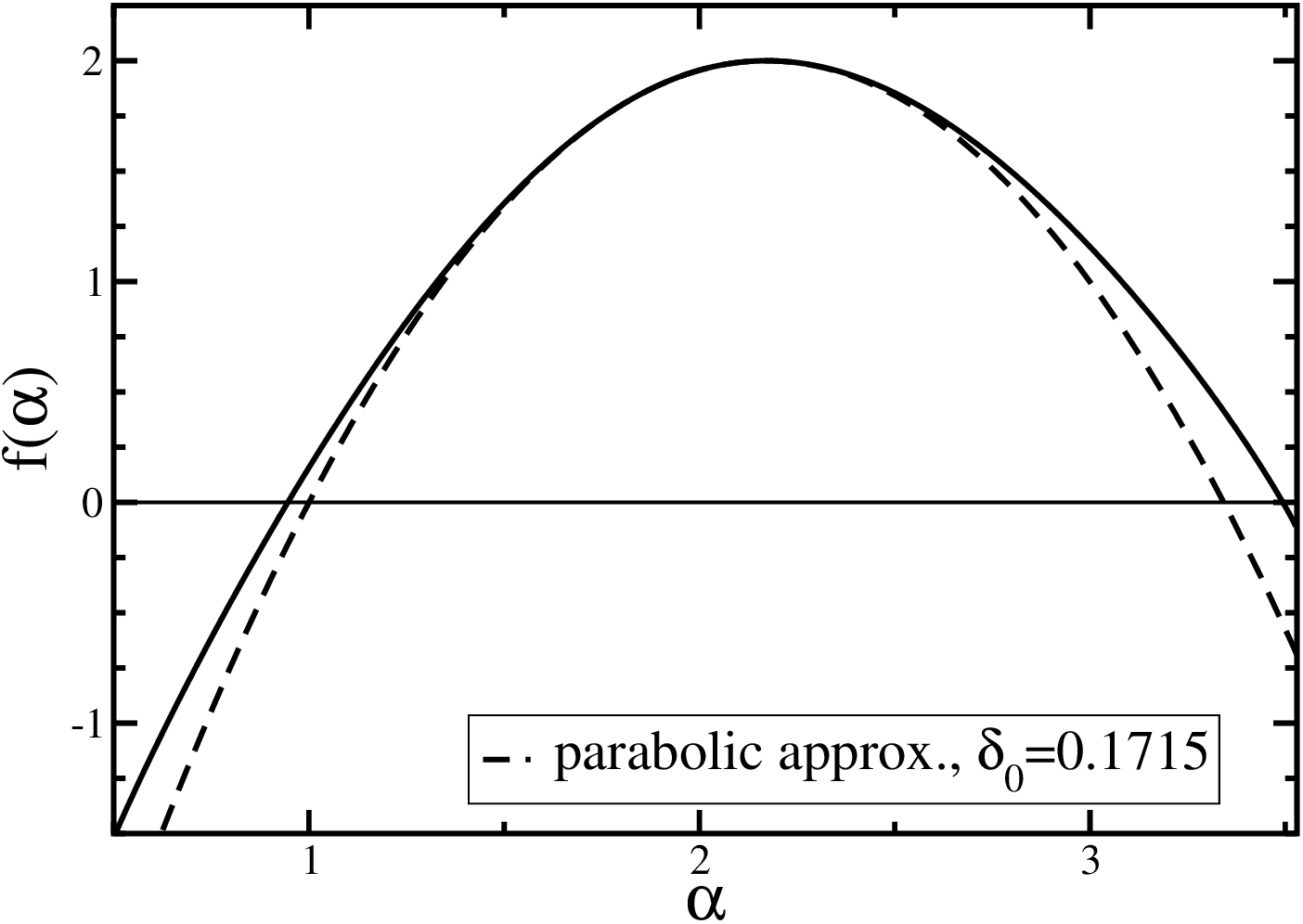}
\caption{$f(\alpha)$ spectrum from data of Fig. \ref{f1}, SU(2) model.
  $f(\alpha)$ is slightly asymmetric and not a 
  parabolic function, which would have meant 
  $f(\alpha){=}2-(\alpha-2-\delta_0)^2/4\delta_0$.}
\label{f3}
\end{figure}
A more conventional object than $\delta_q$ to characterize the
wave function statistics is the Legendre transformed
$
   f(\alpha) = q\alpha - \tau_q,
$
$\alpha_q{=}\partial\tau /\partial q$, displayed in
Fig. \ref{f3}. Even though we have obtained
$\tau_q$ only for $q{\apprge}-1/2$ and therefore are restricted to
$\alpha{\apprle}\alpha_{1/2}$, the spectrum can be reconstructed
also at values $\alpha{\apprge}\alpha_{1/2}$ 
by making use of Eq. (\ref{e3}). \cite{mirlin06a}
Then deviations from parabolicity obtrude. 

\paragraph*{Summary:}
The multifractal spectrum of wave functions at the
2D symplectic Anderson transition has been calculated in the
Ando and SU(2) models with high precision. On a qualitative level, our
results demonstrate that the critical
fixed point is conformally invariant with a nonparabolic
spectrum $\tau_q$. Furthermore, $\delta_q{=}\delta_{1-q}$, as 
predicted from calculations within the nonlinear $\sigma$ model
and thus supports the universality hypothesis.  

We thank L. Schweitzer and K. Yakubo for useful correspondence and
A.~D. Mirlin for valuable discussions and suggestions on the manuscript. 
While finalizing the manuscript, we learned about a closely related
project, with partly overlapping results. \cite{gruzberg06b}

\bibliography{tools.bib}

\begin{thebibliography}{31}
\bibitem{scalingprocedure} 
We employ a standard procedure\cite{zharakeshev95,schweitzer95}
and consider the scaling of
the second moment of the normalized level spacing distribution
$
  J(L,W) = \frac{1}{2} \int_0^\infty \!\! ds \; s^2 P_{L,W}(s) ,
$
where $P_{L,W}(s)$ denotes the probability to find an energy difference
$s$ between two consecutive eigenvalues. 
Our renormalization group ansatz incorporating irrelevant scaling fields
is
$
  J(L,W) = \tilde{J}_{R} (\phi L^{1/\nu})
                + \chi L^{-y} \tilde{J}_{I} (\phi L^{1/\nu}) + \ldots
$
with $\phi(W){=} |W{-}W_c|/W_c$ ($\chi(W)$) being the relevant (irrelevant)
variables.
Results of the scaling analysis are $W_c{=}5.85{\pm}0.025$,
$\nu{=}2.74\pm0.12$, and $y{=}1.5{\pm}0.5$.
The large error of the
exponents result from the uncertainty in $W_c$.
\bibitem{extracttauq}
In recent papers\cite{obuse04}, 
the width $\sigma(L)$ of the distribution function
$\mathcal{P}(\tau_2)$ of
$\tau_2$ for different system sizes is investigated. A nonzero value in the
limit of large system sizes is found, if an approximation scheme is
used that assumes power law corrections to scaling, $L^{-|y|}$.
However, in reality finite size 
corrections to $\sigma(L)$ are logarithmic rather than power
law. Indeed, the
published data ([\onlinecite{obuse04}], Fig. 6) are consistent with
a slow, logarithmic flow of $\sigma(L)$ to zero in agreement with 
the self averaging nature of the multifractal exponents $\tau_q$ and
in contrast to earlier claims made in the literature\cite{parshin99}.
\end{thebibliography}

\end{document}